\documentstyle[11pt]{article}
\begin{document}
\large
\begin{center}
{\Large
{\bf ON GAUGE FIXING IN THE LAGRANGIAN FORMALISM OF
SUPERFIELD BRST QUANTIZATION\\}}

\vspace{.5cm}

Bodo Geyer\footnote{E-mail: geyer@itp.uni-leipzig.de}\\
{\normalsize\it Center of Theoretical Studies,
Leipzig University,\\
Augustusplatz 10/11, 04109 Leipzig, Germany}

\vspace{.5cm}

Petr M. Lavrov\footnote{E-mail: lavrov@tspu.edu.ru} and Pavel
Yu. Moshin \\ {\normalsize\it Tomsk State Pedagogical
University, 634041 Tomsk, Russia}\\
\end{center}

\vspace{.5cm}

\begin{quotation}
\normalsize
\setlength{\baselineskip}{15pt}
\noindent
We propose a modification of the gauge-fixing procedure in the Lagrangian
method of superfield BRST quantization for general gauge theories, which
simultaneously provides a natural generalization of the well-known BV
quantization scheme as far as gauge-fixing is concerned. A superfield form
of BRST symmetry for the vacuum functional is found. The
gauge-in\-de\-pen\-dence of the S-matrix is established.
\end{quotation}

\vspace{.5cm}

\section{Introduction}

Recently, there has been a fairly large amount of papers \cite{BT,HSV,JM,BD,LMR}
devoted to various superfield extensions of the BV quantization method \cite{BV}
for gauge theories. Thus, in \cite{BT} a geometric representation of BRST
transformations \cite{BRST} in the form of supertranslations in superspace was
realized; in \cite{HSV,JM} a superspace formulation of the action and BRST
transformations for Yang-Mills theories was found; in \cite{BD} a superfield
representation of the generating operator $\Delta$ in the BV method was suggested;
in \cite{LMR} a closed superfield form of the BV quantization method \cite{BV}
was obtained. In the study of \cite{BnTn}, a multilevel generalization of the BV
quantization formalism has been proposed, which ensures an invariant description
of field-antifelds variables.

It is well-known that performing the quantization of a gauge theory is
neccessarily related to introducing a gauge-fixing procedure needed for
constructing the Green functions and the S-matrix. The methods \cite{LMR,BV}
implement gauge-fixing underlied by an appropriate choice of the
{\em fermionic} functional, with the only restriction being the
non-degeneracy of the corresponding generating functional.

In the framework of the multilevel generalization \cite{BnTn} of the BV
formalism, gauge-fixing was introduced by means of hypergauge functions
depending on the entire set of field-antifield variables. The hypergauge
functions enter the quantum action in {\em linear} combinations with the
corresponding Lagrange multipliers, and are subject to certain involution
relations expressed in terms of the same antibracket operation that occurs
in the generating equation determining the quantum action. These involution
relations serve to ensure the BRST invariance of the vacuum functional, thus
providing effectively for its independence from hypergauge variations of
canonical form in the antibracket sense.

In this paper we consider another generalization of the BV quantization
scheme obtained by modifying the superfield formalism \cite{LMR} in such
a way that the gauge is now introduced with the help of a gauge-fixing
{\em bosonic} functional which depends generally on the entire set of
variables of the formalism, including sources and Lagrange multipliers,
and which is subject to a generating equation formally analogous to the
equation determining the quantum action. On the one hand, this approach
to gauge-fixing guarantees the independence of the vacuum functional
(and, hence, that of the S-matrix) from any particular choice of the
gauge. On the other hand, within this framework, as compared to that
of \cite{BnTn}, we are no longer confined to linear dependence on the
fields of Lagrange multipliers. Remarkably, the generating equation
introducing the gauge admits of a solution which is identical with the
gauge-fixing functional found in the original version of \cite{LMR}, and
which, in particular, includes the well-known gauge-fixing condition of
the BV formalism.

Our starting point at the classical level is a general gauge theory with
the well-known structure of the configuration space $\phi^A,
\varepsilon(\phi^A)=\varepsilon_A$, described by the rules \cite{BV},
depending on whether the theory is a reducible or irreducible one.

We use the condensed notations \cite{DeW} and the conventions adopted in
\cite{LMR}.

\section{Modified Superfield BRST Quantization}

In this section we shall extend the procedure \cite{LMR} of superfield
quantization underlied by the principle of BRST invariance. We first
introduce a superspace $(x^\mu,\theta)$ spanned by space-time
coordinates $x^\mu$, $\mu=(0,1,\ldots,D-1)$, and a scalar anticommuting
coordinate $\theta$. Let $\Phi^A(\theta)$ be a set of superfields,
which is accompanied by a set of the corresponding
super-antifields $\Phi^*_A(\theta)$ with the Grassmann parities
$\varepsilon(\Phi^A)\equiv\varepsilon_A,\,
\varepsilon(\Phi^*_A)=\varepsilon_A+1$, and is subjected to the boundary
condition
\begin{eqnarray}
\label{BondSQ}
 \Phi^A(\theta)|_{\theta =0}=\phi^A.
\end{eqnarray}
We define the vacuum functional $Z$ as the following functional integral:
\begin{eqnarray}
\label{ZSQ}
Z=\int
 d\Phi\;d\Phi^*\rho[\Phi^*]
\exp\bigg\{\frac{i}{\hbar}\bigg(W[\Phi,\Phi^*] +
X[\Phi,\Phi^*] +\Phi^*\Phi\bigg)\bigg\}.
\end{eqnarray}
Here, $W=W[\Phi,\Phi^*]$ is the quantum action which obeys the
generating equation
\begin{eqnarray}
\label{GEq1SQ}
\frac{1}{2}(W,W)+VW=i\hbar\Delta W,
\end{eqnarray}
while the (bosonic) gauge-fixing functional $X=X[\Phi,\Phi^*]$
is required to satisfy the equation
\begin{eqnarray}
\label{GEqXSQ}
\frac{1}{2}(X,X)-UX=i\hbar\Delta X.
\end{eqnarray}

In Eqs.~(\ref{GEq1SQ}), (\ref{GEqXSQ}),
we have used the antibracket
$(\;,\;)$ expressed in terms of arbitrary functionals $F=F[\Phi,\Phi^*]$,
$G=G[\Phi,\Phi^*]$ by the rule \cite{LMR}
\begin{eqnarray}
\label{ABSQ}
 (F,G)&=&\int d\theta\bigg\{\frac{\delta F}{\delta\Phi^A(\theta)}
 \frac{\partial}{\partial\theta}\frac{\delta
 G}{\delta\Phi^*_A(\theta)}(-1)^{\varepsilon_A+1}
-(-1)^{(\varepsilon(F)+1)(\varepsilon(G)+1)}(F\leftrightarrow
 G)\bigg\}.
\end{eqnarray}
 We have also used the operators $\Delta$, $U$ and $V$
\begin{eqnarray}
\label{DeltaSQ}
\Delta=-\int d\theta(-1)^{\varepsilon_A}
\frac{\delta_l} {\delta\Phi^A(\theta)}
 \frac{\partial}{\partial\theta}\frac{\delta}
 {\delta\Phi^*_A(\theta)},
\end{eqnarray}
\begin{eqnarray}
\label{VSQ}
 V=-\int d\theta\frac{\partial\Phi^*_A(\theta)}{\partial\theta}
 \frac{\delta}{\delta\Phi^*_A(\theta)},
\end{eqnarray}
\begin{eqnarray}
\label{USQ}
 U=-\int d\theta \frac{\partial\Phi^A(\theta)}
 {\partial\theta}\frac{\delta_l}{\delta\Phi^A(\theta)}
\end{eqnarray}
 (derivatives with respect to $\theta$ are understood as acting from
 the left) as well as the functionals $\rho[\Phi^*]$ and
 $\Phi^*\Phi$, i.e.
\begin{eqnarray}
\label{WFuncSQ}
\rho[\Phi^*]&=&\delta\bigg(\int
 d\theta\Phi^*(\theta)\bigg),
\end{eqnarray}
\begin{eqnarray}
\Phi^*\Phi&=&\int
 d\theta\Phi^*_A(\theta)\Phi^A(\theta).
\end{eqnarray}

The algebraic properties of both the antibracket (\ref{ABSQ}) and the
operators (\ref{DeltaSQ}), (\ref{VSQ}), (\ref{USQ}) were studied in detail
in \cite{LMR}. The algebra of the above operators has the form
\begin{eqnarray}
\label{NilSQ}
{\Delta}^2=0,\;\; U^2=0, \;\;V^2=0,
\end{eqnarray}
\begin{eqnarray}
\label{AlgSQ}
UV+VU=0,\;\;
\Delta U+U\Delta=0,\;\;
\Delta V+V\Delta=0.
\end{eqnarray}

It is convenient to recast the equations (\ref{GEq1SQ}), (\ref{GEqXSQ}) into
the equivalent form
\begin{eqnarray}
\label{GEq2SQ}
\bar{\Delta}\exp\bigg\{\frac{i}{\hbar}W\bigg\}=0,
\end{eqnarray}
\begin{eqnarray}
\label{GEqX1SQ}
\tilde{\Delta}\exp\bigg\{\frac{i}{\hbar}X\bigg\}=0,
\end{eqnarray}
using the operators
\begin{eqnarray}
\label{Deltas_nSQ}
\bar{\Delta}=\Delta + \frac{i}{\hbar}V,\;\;
\tilde{\Delta}=\Delta - \frac{i}{\hbar}U,
\end{eqnarray}
whose algebra, by virtue of the properties (\ref{NilSQ}), (\ref{AlgSQ}),
reads as follows:
\begin{eqnarray}
\label{Alg_nSQ}
{\bar{\Delta}}^2=0,\;\; {\tilde{\Delta}}^2=0,\;\;
\bar{\Delta}\tilde{\Delta} + \tilde{\Delta}\bar{\Delta}=0.
\end{eqnarray}

Using the nilpotency of the operator $U$, we observe that any functional
$X=U\Psi[\Phi]$, with $\Psi[\Phi]$ being an arbitrary fermionic functional,
is obviously a
solution of Eq.~(\ref{GEqXSQ}). The above expression gives the
precise form of the gauge-fixing functional proposed by the study of
\cite{LMR} when formulating the rules of superfield BRST quantization.

A remarkable property of the integrand in (\ref{ZSQ}) is its invariance
under the following transformations of global supersymmetry with an
anticommuting parameter $\mu$:
\begin{eqnarray}
\label{BRSTSQ}
&& \delta\Phi^A(\theta)=\mu U\Phi^A(\theta) + (\Phi^A(\theta), X
- W)\mu,
\nonumber
\\ &&
\delta\Phi^*_A(\theta)=\mu V\Phi^*_A(\theta) +
 (\Phi^*_A(\theta), X - W)\mu.
\end{eqnarray}

Here, we have taken into account, first, the expressions of the derivatives
\begin{eqnarray}
\label{}
\nonumber
\frac{\delta_l\Phi^A(\theta)}{\delta\Phi^B(\theta^{'})}
=(-1)^{\epsilon_A} \delta(\theta^{'}- \theta)\delta^A_B
=(-1)^{\epsilon_A}\frac{\delta\Phi^A(\theta)}{\delta\Phi^B(\theta^{'})},
\end{eqnarray}
\begin{eqnarray}
\nonumber
\frac{\delta\Phi^*_A(\theta)}{\delta\Phi^*_B(\theta^{'})}
=(-1)^{\epsilon_A +1} \delta(\theta^{'}- \theta)\delta^B_A,
\end{eqnarray}
following from the definition of integration over the Grassmann
variable $\theta$
\begin{eqnarray}
\label{}
\nonumber
\int d\theta\;\theta = 1, \;\; \int d\theta =0, \;\; F(\theta) =
\int d\theta^{'}\delta(\theta^{'}- \theta) F(\theta^{'}),
\end{eqnarray}
\begin{eqnarray}
\nonumber
\;\; \delta(\theta^{'} -\theta)= \theta^{'} -\theta,
\end{eqnarray}
second, the invariance of the weight functional (\ref{WFuncSQ}) under the
transformations (\ref{BRSTSQ}), $\delta\rho[\Phi^*] = 0$, third, the fact
that under these transformations we have
\begin{eqnarray}
\label{}
\delta (W + X + \Phi^*\Phi) = 2\mu \bigg(\frac{1}{2}(W,W) + VW -
\frac{1}{2}(X,X) + UX\bigg),
\end{eqnarray}
and, finally, the fact that the corresponding Jacobian $Y$ has the form
\begin{eqnarray}
\label{}
Y = \exp (2\mu\Delta W - 2\mu\Delta X).
\end{eqnarray}
Eqs. (\ref{BRSTSQ}) are the transformations of BRST symmetry in the framework
of superfield quantization based on the gauge-fixing functional
$X$ introduced as a solution
of the cor\-res\-pon\-ding generating equation (\ref{GEqXSQ}).

We now consider the gauge-dependence of the vacuum functional
$Z$, Eq.~(\ref{ZSQ}). Note, in the first place, that any admissible variation
$\delta X$ of the gauge-fixing functional $X$ should satisfy the
equation
\begin{eqnarray}
\label{}
(X,\delta X) - U\delta X = i\hbar \Delta \delta X,
\end{eqnarray}
which can be represented in the form
\begin{eqnarray}
\label{VarXSQ}
 \widehat{Q}(X)\delta X=0.
\end{eqnarray}
Here, we have introduced the graded linear, nilpotent operator
$\hat{Q}(X)$,
\begin{eqnarray}
\label{QSQ}
 \widehat{Q}(X)=\widehat{\cal B}(X)-i\hbar\tilde{\Delta}, \quad
 \widehat{Q}^2(X) = 0,
\end{eqnarray}
where $\widehat{\cal B}(X)$ stands for an operator acting by the rule
\begin{eqnarray}
\label{}
 (X,F)\equiv\widehat{\cal B}(X)F,
\end{eqnarray}
and possessing the property
\begin{eqnarray}
\label{}
 \widehat{\cal B}^2(X)=\widehat{\cal B}\left(\frac{1}{2}(X,X)\right).
\end{eqnarray}
By the nilpotency of the operator $\widehat{Q}(X)$, any functional of the form
\begin{eqnarray}
\label{DeltaXSQ}
 \delta X=\widehat{Q}(X)\delta \Psi,
\end{eqnarray}
with $\delta \Psi$ being an arbitrary fermionic functional, obeys the
equation (\ref{VarXSQ}). Furthermore, as in the theorems proved by the study
of \cite{BLT-L}, one can establish the fact that any solution $\delta X$
of Eq.~(\ref{VarXSQ}), vanishing when all the variables entering $\delta X$
are equal to zero, has the form (\ref{DeltaXSQ}), with a certain fermionic
functional $\delta \Psi$.

Let $Z_X\equiv Z$ be the value of the vacuum functional (\ref{ZSQ})
related to the gauge condition chosen as a functional
$X$. In the vacuum functional $Z_{X+\delta X}$ we now make the change
of variables (\ref{BRSTSQ}) with a functional
$\mu=\mu[\Phi,\Phi^*]$, accompanied by an additional change
\begin{eqnarray}
\label{}
 \delta\Phi^A=(\Phi^A,\delta Y),\quad\delta\Phi^*_A=(\Phi^*_A,\delta
Y),\quad\varepsilon(\delta Y)=1,
\end{eqnarray}
where $\delta Y=-i\hbar\mu[\Phi,\Phi^*]$. We obtain
\begin{eqnarray}
\label{ZX+SQ}
 Z_{X+\delta X}=\int d\Phi\,d\Phi^* \rho[\Phi^*]
 \exp\left\{\frac{i}{\hbar}\bigg(W+X+\delta X+\delta X_1
 +\Phi^*\Phi\bigg)\right\}.
\end{eqnarray}
In (\ref{ZX+SQ}), we have denoted
\begin{eqnarray}
\label{}
 \delta X_1=2\bigg((X,\delta Y)-U\delta Y-i\hbar\Delta\delta
 Y\bigg)=2\widehat{Q}(X)\delta Y.
\end{eqnarray}
Let the functional $\delta Y$ be chosen in the form  (recall that
$\delta X=\widehat{Q}(X)\delta \Psi$)
\begin{eqnarray}
\label{}
 \delta Y=-\frac{1}{2}\delta\Psi.
\end{eqnarray}
Thereby we find
\begin{eqnarray}
\label{GIndZSQ}
 Z_{X+\delta X}=Z_X,
\end{eqnarray}
which implies the fact that the vacuum functional (and, hence, the S-matrix,
by the equivalence theorem \cite{KT}) does not depend on the gauge.

\section{Discussion}

In the previous section we have extended the superfield quantization method
\cite{LMR} for general gauge theories to encompass the concept of
gauge-fixing introduced by means of an appropriate generating equation
(see (\ref{GEqXSQ})).

The method of the modified superfield BRST quantization permits one to generalize
the BV quantization scheme \cite{BV} as far as the procedure of gauge-fixing
is concerned. In fact, consider the component representation of the
superfields $\Phi^A(\theta)$ and super-antifields $\Phi^*_A(\theta)$
\begin{eqnarray}
\nonumber
 \Phi^A(\theta)= \phi^A + \lambda^A\theta,\quad
 \Phi^*_A(\theta)=\phi^*_A - \theta J_A, \\
\nonumber
 \varepsilon(\phi^A)= \varepsilon(J_A)= \varepsilon_A, \quad
 \varepsilon(\phi^*_A)= \varepsilon(\lambda^A)= \varepsilon_A + 1.
\end{eqnarray}
The set of variables $\phi^A, \phi^*_A, \lambda^A, J_A$ is identical with
the complete set of variables of the BV method \cite{BV}.

Expressed in the component form, the antibracket (\ref{ABSQ})
and the operator $\Delta$ (\ref{DeltaSQ}) coincide with the corresponding
objects of the BV method \cite{BV}
\begin{eqnarray}
\nonumber
 (F,G)=\frac{\delta F}{\delta\phi^A} \frac{\delta G}{\delta \phi^*_A}
 -(-1)^{(\varepsilon(F)+1)(\varepsilon(G)+1)}(F\leftrightarrow G)\;,
\end{eqnarray}
\begin{eqnarray}
\nonumber
 \Delta &=& (-1)^{\varepsilon_A}\frac{\delta_l}{\delta \phi^A} \frac
 {\delta}{\delta \phi^*_A}\;.
\end{eqnarray}

Representing the integration measure in terms of the components
\begin{eqnarray}
\nonumber
 d\Phi\;d\Phi^*\;\rho[\Phi^*]=d\phi\;d\phi^*\;d\lambda\;dJ\;\delta(J),
\end{eqnarray}
we observe that solutions of the generating equations determining
the action $W$ when $J_A=0$ may be sought among solutions of the master
equation applied by the BV method
\begin{eqnarray}
\label{BVgeneq}
\frac{1}{2}(W,W)=i\hbar\Delta W,
\end{eqnarray}
since the operator $V$ (\ref{VSQ})
\begin{eqnarray}
\nonumber
 V=-J_A \frac{\delta}{\delta \phi^*_A}
\end{eqnarray}
vanishes when $J_A=0$.
Restricting ourselves to functionals $W$ independent of $\lambda^A$,
and taking into account the component form of $\Phi^*\Phi$, i.e.
\begin{eqnarray}
\nonumber
\Phi^* \Phi = \phi^*_A \lambda^A -
 J_A\phi^A,
\end{eqnarray}
we arrive at the following representation of the vacuum functional in
Eq.~(\ref{ZSQ}):
\begin{eqnarray}
\label{GFmodBV}
 Z=\int
 d\phi\;d\phi^*\;d\lambda\exp\bigg\{\frac{i}{\hbar}\bigg[W(\phi,\phi^*)
 + X(\phi,\phi^*,\lambda) + \phi^*_A \lambda^A\bigg]\bigg\}.
\end{eqnarray}

The above result may be considered as an extention of the BV quantization
procedure \cite{BV} to a more general case of gauge-fixing. In fact,
as stated above, the functional $X=U\Psi[\Phi]$ is a solution of the
generating equation (\ref{GEqXSQ}). From the component representation of the
operator $U$
\begin{eqnarray}
\nonumber
 U=-(-1)^{\varepsilon_A}\lambda^A\frac{\delta_l}{\delta \phi^A}\;,
\end{eqnarray}
provided the functional $\Psi$ is independent of the fields
$\lambda^A$, $\Psi=\Psi(\phi)$, it follows that the gauge-fixing functional
$X$
\begin{eqnarray}
\nonumber
 X(\phi,\lambda)=-\frac{\delta\Psi(\phi)}{\delta \phi^A}\lambda^A
\end{eqnarray}
becomes identical with the gauge applied by the BV quantization method.

It is instructive to compare the general framework of the present study with
that of the multilevel formalism \cite{BnTn}. Notice that in writing down the
relations of \cite{BnTn} we apply the notations of the present paper, thereby
assuming an explicit separation of field-antifield variables in terms of
the Darboux coordinates.

If one assumes the measure density of the functional integration to be
trivial, the general ansatz for the first-level vacuum functional
\cite{BnTn} reads as follows:
\begin{eqnarray}
\label{BnTn_Z}
 Z=\int
 d\phi\;d\phi^*\;d\lambda\exp\bigg\{\frac{i}{\hbar}\bigg[W(\phi,\phi^*)
 + G_A(\phi,\phi^*)\lambda^A\bigg]\bigg\},
\end{eqnarray}
where $W(\phi,\phi^*)$ is a quantum action subject to the generating
equations (\ref{BVgeneq}), while $G_A(\phi,\phi^*)$ are gauge-fixing
functions, or hypergauge functions, according to the terminology of
\cite{BnTn}.

The hypergauge functions $G_A=G_A(\phi,\phi^*)$ are assumed \cite{BnTn} to
satisfy the involution relations
\begin{equation}
\label{BnTn_inv}
 (G_A,G_B)=G_CU^C_{AB},
\end{equation}
accompanied by the so-called unimodularity conditions
\begin{equation}
\label{BnTn_unim}
 \Delta G_A-U^B_{BA}(-1)^{\varepsilon_B}=G_BV^B_{A},\;\;\;
 V^A_A=G_A\tilde{G}^A
\end{equation}
with certain structure functions $U^C_{AB}$, $V^B_A$, $\tilde{G}^A$.

As shown in \cite{BnTn}, the set of the conditions (\ref{BnTn_inv}),
(\ref{BnTn_unim}), combined with the generating equation (\ref{BVgeneq}),
serves to ensure the invariance of the vacuum functional (\ref{BnTn_Z})
under the BRST transformations
\[
 \delta\phi^A=(\phi^A,W-G_B\lambda^B)\mu,\;\;\;
 \delta\phi^*_A=(\phi^*_A,W-G_B\lambda^B)\mu,\;\;\;
\]
\begin{equation}
 \label{BnTn_tr}
 \delta\lambda^A=\left(U^A_{BC}\lambda^C\lambda^B(-1)^{\varepsilon_B}
 -2i\hbar V^A_B\lambda^B-2(i\hbar)^2\tilde{G}^A\right)\mu,
\end{equation}
which, in their turn, provide for the independence of the vacuum functional
from hypergauge variations of canonical form
\begin{eqnarray}
 \label{BnTn_gvar}
 \delta G_A=(G_A,\delta Y),\;\;\;\varepsilon(\delta Y)=1.
\end{eqnarray}
It should be noted that the above variation (\ref{BnTn_gvar}) is compatible
with the conditions (\ref{BnTn_inv}), (\ref{BnTn_unim}). Within the
assumption that the hypergauge functions are solvable with respect to the
antifields, the first-level vacuum functional (\ref{BnTn_Z}) reduces to the
well-known form \cite{BV} of the BV formalism.

In view of the above, we now return to the vacuum functional (\ref{GFmodBV})
of the modified superfield formalism and consider the particular case of
linear dependence of the gauge-fixing functional $X(\phi,\phi^*,\lambda)$ on
the Lagrange multipliers, i.e.
\[
 X(\phi,\phi^*,\lambda)+\phi^*_A\lambda^A\equiv G_A(\phi,\phi^*)\lambda^A.
\]
Then, firstly, equation (\ref{GEqXSQ}) determining $X$ reduces to the
conditions
\begin{eqnarray}
 \label{lin_eq}
 (G_A,G_B)=0,\;\;\;\Delta G_A=0,
\end{eqnarray}
and, secondly, the transformations of BRST symmetry for the vacuum
functional (\ref{GFmodBV}) take on the form
\begin{eqnarray}
 \label{trans}
 \delta\phi^A=(\phi^A,G_B\lambda^B-W)\mu,\;\;
 \delta\phi^*_A=(\phi^*_A,G_B\lambda^B-W)\mu,\;\;\delta\lambda^A=0,
\end{eqnarray}
which follows from (\ref{BRSTSQ}) when $J_A=0$.

Obviously, equations (\ref{lin_eq}), are identical with (\ref{BnTn_inv}),
(\ref{BnTn_unim}) in the particular case of vanishing structure
functions $U^C_{AB}$, $V^B_A$, $\tilde{G}^A$, whereas the transformations
(\ref{trans}), similarly, present a particular case of (\ref{BnTn_tr}).

Note also that there is a close connection between the result (\ref{GFmodBV}),
obtained as a by-product of our generalization of the superfield quantization
\cite{LMR}, and a generalization of the vacuum functional of the
BV-formalism proposed by the study of \cite{BBD}. In \cite{BBD} it was shown
that the vacuum functional
\begin{eqnarray}
\label{GFBBD}
 Z=\int
 d\phi\;d\phi^*\;d\lambda\exp\bigg\{\frac{i}{\hbar}\bigg[W(\phi,\phi^*)
 + X(\phi,\phi^*,\lambda)\bigg]\bigg\},
\end{eqnarray}
with both functionals $W$ and $X$ subject to the master equation of the
BV-formalism
\[
 \exp\{(i/\hbar)W\}=0,\;\;\;\exp\{(i/\hbar)X\}=0,
\]
does not depend on the choice of the gauge functional. It is clear that the
functional $X'=X+\phi^*_A\lambda^A$ in (\ref{GFmodBV}) obeys the master
equation $\exp\{(i/\hbar)X'\}=0$, and therefore, in terms of $X'$, the
functional (\ref{GFmodBV}) coincides with the one proposed by \cite{BBD}.

Summarizing, both the study of the present paper and that of \cite{BnTn}
apply gauge-fixing conditions depending on the whole set of field-antifield
variables; they both contain the BV method as a particular case and ensure
the gauge-independence of the S-matrix. At the same time, in the present
approach the gauge-independence is encoded in BRST transformations controlled
by generating equations imposed on the entire gauge part of the quantum
action, whereas in the framework of \cite{BnTn} this role is played by
unimoduar involution relations imposed on hypergauge functions.
Despite the formal similarity between the vacuum functional (\ref{GFmodBV})
proposed by the present study and the ansatz (\ref{BnTn_Z}) suggested by
\cite{BnTn} the two methods appear to be independent from each other,
being different generalizations of the BV method. However, as is seen
from the above comparison, the vacuum functional (\ref{GFmodBV}) proposed
by the present study becomes identical, in the particular case of linear
dependence on the Lagrange multipliers, with the first-level vacuum
functional \cite{BnTn} considered in the case of a trivial integration
measure and vanishing structure functions. On the other hand, the
vacuum functional (\ref{GFmodBV}) can be regarded as a formal extension
of (\ref{BnTn_Z}) in the sense that it admits of more than linear
dependence on the Lagrange multipliers.

\section*{Acknowledgments}

The work of P.M.L. and P.Yu.M. was supported in part by a grant of the
Ministry of General and Professional Education of the Russian Federation
in the field of fundamental sciences, as well as by grant RFBR
99-02-16617. The work of P.M.L. was also partially supported by grants
INTAS 96-0308, RFBR-DFG 96-02-00180 and a grant of Saxonian Ministry of
Science and Arts.

\end{document}